\documentclass
[12pt]{article}
\usepackage[dvips]{color}
\usepackage{epsfig}
\usepackage{amsmath}
\usepackage{graphicx}
\begin{document}
\title {Decay widths of large-spin mesons from the non-critical string/gauge duality }
\author{J. Sadeghi $^{a,b}$\thanks{Email:
pouriya@ipm.ir}\hspace{1mm} and S. Heshmatian
$^{a}$\thanks{Email: s.heshmatian@umz.ac.ir}\\
$^a$ {\small {\em  Sciences Faculty, Department of Physics, Mazandaran University,}}\\
{\small {\em P. O. Box 47415-416, Babolsar, Iran}}\\
$^b$ {\small {\em  Institute for Studies in Theoretical Physics and Mathematics (IPM)}}\\
{\small {\em P. O. Box 19395-5531, Tehran, Iran}}}
\maketitle
\begin{abstract}
In this paper, we use the non-critical string/gauge duality to calculate
the decay widths of large-spin mesons. Since it is believed that the
string theory of QCD is not a ten dimensional theory, we expect that the
non-critical versions of ten dimensional black hole backgrounds lead to
better results than the critical ones. For this purpose we concentrate
on the confining theories and consider two different six dimensional black hole
backgrounds. We choose the near extremal $ AdS_6$ model and the near extremal KM model
to compute the decay widths of large-spin mesons. Then, we
present our results from these two non-critical backgrounds and compare them together
with those from the critical models and experimental data.
\end{abstract}

\section{Introduction}
The gauge/gravity duality demonstrates the correspondence between a
gravitational theory in anti de-sitter space and a
gauge theory at large $N$ limit [1-6]. An example for this
correspondence is the relation between type IIB string theory in
ten dimensional  background and $\cal{N}$=4 supersymmetric Yang-Mills
theory on four dimensional boundary of $AdS_5$.  In recent years, using this
correspondence as a powerful tool to study the QCD, has been increased and
lots of papers have been published in this context. For example, the dynamics
of moving quark in a strongly coupled plasma [7-16] and the
jet-quenching parameter [16-24] have been investigated. In
addition, the motion of a quark-antiquark pair in the quark- gluon plasma
has been studied in [25-31].\\
The calculations of decay widths of mesons are so important but they are hard
to do by using the QCD methods because of the strong coupling problems.
So, the holographic methods could help to overcome these difficulties.
Recently, models including the various brane configurations have been
introduced in critical dimensions to describe hadrons in the confining backgrounds.
The model introduced in [32] is one example with $D4/D6$ brane configuration
which leads to the heavy scaler and pseudo-scaler mesons.
Also the model proposed by Sakai and Sugimoto [33] with $D4/D8/\overline{D8}$
in ten dimensional background leads to a nice description of hadron physics. Many
authors have used the holographic methods to study the hadron physics [34-36].
The decay process of mesons is a remarkable task which has been studied
by using the $SS$ model. The authors have calculated the meson masses
and the decay rates by consideration of low-spin meson as small fluctuations of
flavor branes. Of course these models can be used only for low-spin mesons
and can not describe the large-spin mesons anymore.\\
Large spin mesons are interesting because of their phenomenological features,
therefore some authors have chosen the dual string theory description to
study their decay processes in critical dimentions [37]. In this paper
an interesting setup has been proposed to compute the decay widths of mesons.
They have used a semi-classical U-shaped spinning string configuration.
This string can decay into some outgoing mesons by touching one or more
of the flavor branes, splitting and then getting reconnected to the brane due to the quantum
fluctuations. The idea in this paper is to focus on the near wall geometry and build the
string wave function near in this geometry by semi-classical quantization. The authors compared their results
with the Casher-Neuberger-Nussinov model where quark-antiquark pair are connected by a chromoelectric flux tube [42].\\
There is also an old model called "Lund model" describing the mesons decay [38].
There are improvements for this model in the literature where two massive quarks
are connected together by a massless relativistic string [39]. The resulting formula
leads to a better description for the decay widths. It shows that for a decay width
linear in length, the ratio of $\Gamma / M$ is not a constant anymore.
Also the decay process of the open strings [40] and closed strings [41] is studied before by using different methods.\\
The decay widths of both low-spin and large-spin mesons has been studied in critical dimensions [33,37].
Also some calculations for the low-spin mesons
has been done before by using the non-critical string/gauge duality [43].
But the decay widths of high spin mesons has not been studied in the
context of the non-critical duals mesons yet. In holographic QCD,
there is an idea that the string theory in dimensions less
than ten is a good candidate to study the QCD. So, this motivate us
to study the decay widths of large-spin mesons by using the non-critical
version of ten dimensional black hole backgrounds [43-49]. For this purpose, we consider
two different six dimensional backgrounds. The first one is the near extremal
flavored $AdS_6$ which is dual to a four dimensional low energy
effective gauge theory. Mesons in the IR theory are constructed by
the quarks with a mass of the order of temperature. In this model is based on the
near extremal $D4$ branes background and the $D6$ flavor branes added to this background.
The second one is the same as the Klebanov-Maldacena model called KM model [47] with
flavored $AdS_5\times S^1$ background. In the near-extremal background,
there is a system of $D3$ and uncharged $D5$ branes in six dimensional string theory
and one of the gauge theory flat directions is compact on a thermal circle in order to
break supersymetry [43]. The near extremal solution is dual to the four dimensional theory
at finite temperature without supersymmetry and conformal invariance.
In this paper we use the semi-classical model introduced in ref. [37] and do the same
calculations in the two non-critical dual pictures. For this purpose, we choose the flat space-time
approximation for simplicity. According to ref. [37] we construct the wave function for
the string configuration and use it to compute the decay width. \\
This paper is organized as follows: In section 2 we use the non-critical $AdS_6$
background to write an expression for the meson decay width. In section 3 we
we use another non-critical background, the near extremal KM
model with $AdS_5\times S^1$ black hole, and obtain another equation
for the meson decay width for six dimensional string theory. In the last section, we
present our numerical results and compare them with the previous models and the
experimental data of ref. [50]. Also we use the modified relation between the length of horizontal
part of the string and its mass derived in [39] and obtain the decay widths for two non-critical
backgrounds of sections 2, 3 and compare them with the data.

\section{Decay widths in the near extremal \textbf{$AdS_6$} model}
In this paper we use two different non-critical backgrounds to calculate
the mesons decay widths; the near extremal $AdS_6$ model in this section
and the near extremal Klebanov-Maldacena model with $AdS_5\times S^1$
background in the next section. Also we use the method proposed in ref. [37]
for the critical dimensions to calculate the decay widths.\\
First, we briefly review the model of ref. [37] and then use it to do our calculations.
In this paper,  a semi-classical U-shaped spinning string
configuration ith two massive endpoints on the flavor brane is considered. The string is pulled
toward the infrared wall and also extends along it. This configuration is equivalent
to a high spin meson with massive quarks.\\
The string can decay into some outgoing mesons by touching one or more
of the flavor branes, spliting and then getting reconnected to the branes
due to the quantum fluctuations.
The idea in this paper is to focus on the near wall geometry and build the
string wave function near in this geometry by semi-classical quantization.
The total wave function for a classical $U$-shaped string is [37]
\begin{eqnarray}
\Psi\big[\{ {\mathcal N}_n \} \big] = \prod_n \Psi_n\big[{\mathcal N}_n (X^M)\big] \, .
\end{eqnarray}
where ${\mathcal N}_n(X^M)$ are normal coordinates, $X^M$ are the target space coordinates
and $\Psi_n\big[{\mathcal N}_n (X^M)\big]$ are the wave function of normal modes ${\mathcal N}_n$ .
Due to the quantum fluctuations, the string may touch the flavor brane in one or more
points with the probability given by [37]
\begin{eqnarray}
{\mathcal P}_{\text{fluct}} =
\int'_{\{ {\mathcal N}_n\}} \big|\, \Psi\big[\{{\mathcal N}_n\}\big] \, \big|^2\,,
\end{eqnarray}
and only the configurations with the following condition are being integrated
\begin{eqnarray}
\max\big(U(\sigma)\big) \geq U_B\,.
\end{eqnarray}
The splitting probability for the string to is given by [37]
\begin{eqnarray}
{\cal P}_{\text{split}} := \frac{1}{T_{\text{eff}}}\,\frac{\Gamma_{\text{open}}}{L}\,.
\end{eqnarray}
Using this relation, the total decay width takes this form
\begin{eqnarray}
\Gamma = T_{\text{eff}} {\cal P}_{\text{split}}\,\times\,
   \int'_{\{ {\mathcal N}_n\}} \big|\, \Psi\big[\{{\mathcal N}_n\}\big] \, \big|^2
   \, K\big[\{{\mathcal N}_n\}\big]  \, ,
\end{eqnarray}
where $\, K\big[\{{\mathcal N}_n\}\big]$ is a factor with the dimension of $L$ which
measures the size of string sesments intersect the flavor brane.
Finally, the authors in ref.[37] obtained the approximated decay width as following
\begin{eqnarray}
\Gamma_{\text{approx}}
 = \Big( T_{\text{eff}}\,{\mathcal P}_{\text{split}}\,\times\,
 L\,\times\, \kappa_{\text{max}}\Big)\,\times\, {\cal
 P}_{\text{fluct}}\, ,
\end{eqnarray}
where $\kappa_{\text{max}}$ is dimensionless. The fluctuation probability
is the main thing should be computed for the decay width.
${\cal{P}}_{\text{fluct}}$ .\\
Now we are going to use the procedure proposed in ref. [37] to evaluate the decay
widths. \\
First, we use the model introduced above to calculate the meson decay width
in the near extremal $AdS_6$ black hole background [43]. This background is
constructed by near extremal color $D4$-branes and additional $D4/\overline{D4}$
flavor branes. In order to have a non-supersymmetric gauge theory with massless fundamentals,
$D4$ flavor branes are added to the background which are extended along the Minkowski
directions and stretched along the radial direction. The background metric has the form
\begin{eqnarray}
ds^2_6&=&\left( \frac{U}{R_{AdS}} \right)^2 dx_{1,3}^2
+\left( \frac{R_{AdS}}{U} \right)^2 \frac{dU^2}{f(U)} +
\left( \frac{U}{R_{AdS}} \right)^2 f(U) d\theta^2,
\label{unflavmetr}\\
F_{(6)}&=&Q_c \left( \frac{U}{R_{AdS}} \right)^4 dx_0 \wedge
dx_1\wedge dx_2 \wedge dx_3  \wedge dU \wedge d\theta\\
e^\phi &= &\frac{2\sqrt2}{\sqrt3 Q_c}\,\,,
\quad\qquad R_{AdS}^2=\frac{15}{2},
\end{eqnarray}
where $\phi$ is a constant dilaton, $F_{(6)}$ is the $RR$ six-form field strength and
\begin{eqnarray}
f(U)=1-\left( \frac{U_\Lambda}{U} \right)^5.
\end{eqnarray}
The coordinate $\theta$ should be periodic in order to avoid a conical singularity on the horizon
\begin{eqnarray}
\theta \sim \theta + \delta \theta\,\,,\qquad\qquad
\delta\theta=\frac{4\pi R_{AdS}^2}{5 U_\Lambda}\,\,,\
\end{eqnarray}
where $L_\Lambda\equiv\delta\theta$ is the size of the thermal circle
and should be small in order to have a dual four dimensional low energy effective
gauge theory [32]. The mass scale for this non-critical metric is
\begin{eqnarray}
M_\Lambda=\frac{2\pi}{\delta\theta}=
\frac{5}{2} \frac{ \ U_\Lambda}{R_{AdS}^2}.
\end{eqnarray}
At leading order, the fluctuations of horizontal part of the string on the wall experiences a flat
geometry. First, we use the following coordinate to see when the flat approximation is valid [37]
\begin{eqnarray}
\eta^2 = \frac{U - U_\Lambda}{U_\Lambda}\,.
\end{eqnarray}
Then we expand the metric of equation (7) around $\eta\,=\,0$ to quadratic
order and find the following expression for the  $AdS_6$ metric
\begin{eqnarray}
{\rm d}s^2 \sim \left( \frac{U_\Lambda}{R_{AdS}} \right)^{2} (1 + 2 \eta^2) (
\eta_{\mu\nu} {\rm d} X^\mu {\rm d}X^\nu)  + \frac{4}{5}  R_{AdS}^2
{\rm d} \eta^2 + 5  \left( \frac{U_\Lambda}{R_{AdS}} \right)^{2} \eta^2  {\rm d} \theta^2 .
\end{eqnarray}
Then we consider the following solutions of a rotating string at the IR wall
\begin{eqnarray}
T = L \tau\,, \quad
X^1 = L \sin\tau\sin\sigma\,,\quad
X^2 = L \cos\tau\sin\sigma\,,\quad U=U_\Lambda\,,
\end{eqnarray}
where $L$ is the length of the horizontal part of the string. To quantize the
linearized metric (14) by using the Polyakov formulation.
The Polyakov string action in a curved background is given by
\begin{equation}
S = \frac{1}{2\pi\alpha'} \int\!{\rm d}\tau \int_0^{2\pi\sqrt{\alpha'}}\!{\rm d}\sigma
 \, G_{MN} \left[\dot X^M \dot X^N - X^M{}' X^N{}'\right]\,.
\end{equation}
As explained in ref. [37], the fluctuations along the wall directions are
irrelevant to construct the wave function. So, we only consider the fluctuations
in the $\eta$ and $X^\mu$ directions (transverse to the wall).
By expanding the above action around the solutions of (15) and keeping the quadratic
terms in $\eta$, we find the following action
\begin{eqnarray}
S =
\frac{1}{2\pi\alpha'} \left( \frac{U_\Lambda}{R_{AdS}} \right)^{2}  \int\!{\rm d}\tau{\rm d}\sigma\,
 \frac{4}{5} \frac{R_{AdS}^4}{U_\Lambda^2} \left[ \left( \dot\eta^2 - {\eta'}^2\right)
 - b \,\cos^2(\sigma)
\left( 1 + 2 \eta^2 \right) \,\right]
\nonumber\\
+ \left[ \left(1 + 2 \eta^2\right) \left(\delta \dot{X}^\mu \delta
  \dot{X}^\nu \eta_{\mu\nu}  - \delta X^{\mu '} \delta X^{\nu'}
  \eta_{\mu\nu}\right) \right] \, ,
\end{eqnarray}
where $b$ is a dimensionless parameter as following
\begin{eqnarray}
b \equiv \frac{5}{2}\frac{\pi^2 L^2 U_\Lambda^2}{R_{AdS}^4} \, ,
\end{eqnarray}
which determines the effect of curvature. If $b\ll\,1$, the string fluctuations are
small enough, so we can use the flat space approximation with the following coordinate
\begin{eqnarray}
\eta = \sqrt{\frac{5}{4}} \frac{U_\Lambda}{R_{AdS}^2}\, z\,.
\end{eqnarray}
to write the metric (14) in the conformally flat form.
\begin{eqnarray}
{\rm d}s^2 \sim \left( \frac{U_\Lambda}{R_{AdS}} \right)^{2}
\Big(\eta_{\mu\nu} {\rm d} X^\mu {\rm d}X^\nu
 + {\rm d}z^2 \Big) + 5\left( \frac{U_\Lambda}{R_{AdS}}\right)^{2} \eta^2 {\rm d}\theta^2\,.
\end{eqnarray}
Again by expanding the Polyakov action for the fluctuations in the directions transverse
to the wall and use $T= L \tau$, we find
\begin{eqnarray}
S_{\text{fluct}} = \frac{L}{2 \pi \alpha'_{\text{eff(1)}}} \int\!{\rm d} T {\rm d} \sigma\,
\left[ - (\partial_T z)^2 + \frac{1}{L^2} (\partial_\sigma z)^2 \right] \, ,
\end{eqnarray}
where we have neglected the fluctuations in the directions along the wall.
In the above formula, the string effective coupling for the near extremal $AdS_6$ background is as following
\begin{eqnarray}
\alpha'_{\text{eff(1)}} = \alpha' \left( \frac{R_{AdS}}{U_\Lambda}\right)^{2} \, ,
\end{eqnarray}
which is obtained by using the following equation for the non-critical string stretching close
to the horizon of the $AdS_6$ black hole [43]
\begin{eqnarray}
T_{eff(1)} = \frac{1}{2\pi \alpha'}  \sqrt{g_{00} g_{xx}} |_{wall} = \frac{1}{2\pi \alpha'}\left( \frac{U_\Lambda}{R_{AdS}}\right)^{2}
\end{eqnarray}
Then, by imposing the Dirichlet boundary conditions for the fluctuations $z(\sigma,\tau)$,
one can write the following equation
\begin{eqnarray}
z(\sigma,\tau) = \sum_{n>0} z_n \cos(n\sigma)\,.
\end{eqnarray}
We put this in the action (21) and integrate over $\sigma$ coordinate to obtain
\begin{eqnarray}
S_{\text{fluct}} = \frac{L}{2 \alpha'_{\text{eff(1)}}} \int\!{\rm d} T\,
    \left[  \sum_{n>0}\left( - (\partial_T z_n)^2 +
		\frac{n^2}{L^2}
  z_n^2 \right )\right ]\,.
\end{eqnarray}
for the fluctuation in the $z$ direction. From this equation, we can easily find that
the system is similar to infinite number of linear harmonic oscillators.
The frequencies are $n/L$ and the masses are $L/\alpha'_{\text{eff(1)}}$.
We can see that this result is in the form of the critical setup obtained in ref. [37].
But the important difference is in the expression of $\alpha'_{\text{eff(1)}}$ (equation (22))
which is different in critical and non-critical cases. Also the values of $U_{\Lambda}$ and $R_{AdS}^2$
differ from the corresponding critical values.\\
The wave function in the factorized form is [37]
\begin{eqnarray}
\Psi(\{z_n\},\{ x_n \} )= \Psi_{\text{long}}(\{ x_n \}) \times
                          \Psi_{\text{sphere}}(\{ y_n \}) \times
                          \Psi_\theta(\{ \theta_n \}) \times
                          \Psi_{\text{trans}}(\{ z_n \}) \, .
\end{eqnarray}
where $\Psi_{\text{long}}=\Psi_{\text{sphere}} =\Psi_\theta = 1$  because
only the fluctuations transverse to the wall contribute to the fluctuation probability
and those belong to other directions are being integrated. So the wave function
for the transverse directions is written as [37]
\begin{eqnarray}
\Psi[\{z_n \}] = \prod_{n=1}^\infty \Psi_{0}(z_n)\,.
\end{eqnarray}
From equation (25) we can write the following equation for the wave
functions of the string compared to the harmonic oscillators
\begin{eqnarray}
\Psi_0(z_n) = \left(\frac{n}{\pi \alpha'_{\text{eff(1)}} } \right)^{1/4} \exp\left(-
  \frac{n}{2\alpha'_{\text{eff(1)}}} \, z_n^2\right) \, .
\end{eqnarray}
This equation is also similar to the critical case [37], but the difference is in
the $\alpha'_{\text{eff(1)}}$ equation. All oscillators are in their ground state
and there is no relevant excited mode. Also there is the following condition [37]
\begin{eqnarray}
\sum_{n > 0} \big|z_n\big| \leq z_B\,.
\end{eqnarray}
which means that if one add all the modes constructively, the total amplitude is still smaller than $z_B$.
This condition leads to the following expression for the upper bound on the fluctuation probability [37]
\begin{eqnarray}
{\cal P}^{\text{max}}_{\text{fluct}} =  1 - \idotsint\limits_{\sum_{n>0} |z_n| \leq z_B}\,
\prod_{n=1}^\infty {\rm d}z_n \, \big| \Psi( \{ z_n \}) \big|^2  \, .
\end{eqnarray}
There is also another condition for the string not to touch the brane which leads to
a lower bound for the string fluctuation probability as following [37]
\begin{eqnarray}
{\mathcal P}^{\text{min}}_{\text{fluct}} = 1- \lim_{N \rightarrow \infty} \,
\int_0^{z_B} \! {\rm d}z_1 \int_0^{z_B} \! {\rm d}z_2 \cdots
\int_0^{z_B} \! {\rm d}z_N \, \,  \big| \Psi( \{ z_n \}) \big|^2  \ .
\end{eqnarray}
The authors in ref [37] have evaluated this integral numerically and fitted their
result to a Gaussian with the following expression
\begin{eqnarray}
{\mathcal P}^{\text{min}}_{\text{fluct}} \approx \exp\left( - 1.3\frac{z_B^2}{\alpha'_{\text{eff}}}\right) \, .
\end{eqnarray}
They have obtained this result by using the  ${\mathcal P}^{\text{min}}_{\text{fluct}}$
plot in terms of $z_B/\sqrt{\alpha'_{\text{eff}}}$. If we do the same process, the following
equation for ${\mathcal P}^{\text{min}}_{\text{fluct}}$ is obtained
\begin{eqnarray}
{\mathcal P}^{\text{min}}_{\text{fluct}} \approx \exp\left( - 1.3\frac{z_B^2}{\alpha'_{\text{eff(1)}}}\right) \,
\end{eqnarray}
This equation is the same as ref. [37], except that the $\alpha'_{\text{eff}}$ is replaced by $\alpha'_{\text{eff(1)}}$.
Then we put equation (33) into the equation (6) and find the decay width
of large-spin mesons in the flat space approximation as following
\begin{eqnarray}
\Gamma_{\text{flat}} =  \Big( \text{const}.\times
T_{\text{eff}}\,{\mathcal P}_{\text{split}}\,\times\,
 L \Big) \,\times\, \exp\left( - 1.3\frac{z_B^2}{\alpha'_{\text{eff(1)}}}\right)  \, .
\end{eqnarray}
This is the decay width we obtain by using the near extremal $AdS_6$ black hole background.
The difference between our result and the result of critical dimensions (ref. [37]) is the
precise form of the exponent. Since the string effective coupling is different in these two
backgrounds, this leads to different results for the decay width. Also this difference exists
in the case of the Lund model [38]. We present our numerical results in the last section.\\
Equation (34) shows that the ratio $\Gamma/L$ is constant on
the same Regge trajectory just like the results of the ref. [37] and
the Lund model. But the experimental data do not support this result exactly.
As mentioned in ref. [37], this deviation could be justify by the fact that the Regge
trajectories in the nature are not straight lines and one should
consider the effect of two massive endpoints as following [39]
\begin{eqnarray}
\frac{L}{M}=\frac{2}{\pi\,T_{\text{eff}}}- \frac{m_1+m_2}{2 T_{\text{eff}} M}
+{\cal O}\left(\frac{m_i^2}{M^2}\right)\, .
\end{eqnarray}
In ref. [37], this relation has been applied to the decay rates. The authors showed
that in the case of $\Gamma$ linear in $L$, the $\Gamma/ M$ ratio in not a constant.
They concluded that as $M$ increase, this ratio increases and reaches to a
universal value for large $M$. So, equation (35) leads to a better result for
the decay width which is compatible with the experimental data. We also use this
equation together with equation (22) to find the decay widths for $a$ and $f$ mesons.
We discuss our results in the last section.

\section{Decay widths in the near extremal $KM$ model}
In this section we use another non-critical background to compute the decay width of
large-spin mesons. We choose the near extremal version of Klebanov-Maldacena model, $AdS_5\times S^1$.
The system is composed of $D3$ color branes and uncharged $D5$ flavor branes in six
dimensional non-critical string theory. Here, we consider one of gauge theory flat dimensions
to be compact on a thermal circle. This background is dual to a four dimensional field theory
at finite temperature with fundamental flavors. The $AdS_5\times S^1$ metric is given by [43]
\begin{eqnarray}
ds^2=\left(\frac{U}{R'_{AdS}}\right)^2\left( dx^2_{1,2}+
\left(1-\left(\frac{U_\Lambda}{U}\right)^4\right)d\theta^2\right)
+\left(\frac{R'_{AdS}}{U}\right)^2\frac{dU^2}{1-\left(\frac{U_\Lambda}{U}\right)^4}+R_{S^1}^2 d\varphi^2
\, ,
\end{eqnarray}
with $R'_{AdS}=\sqrt{6}$, $R_{S^1}^2=\frac{4 Q_c^2}{3 Q_f^2}$, $U_\Lambda^4=2 b_1 R_{AdS}^{'4}$ and $e^{\phi_0}=\frac{4}{3Q_f}$.
The RR 5-form field strength is
\begin{eqnarray}
F_5=Q_c\left( \frac{U}{R'_{AdS}}  \right)^3 dx^0\wedge dx^1 \wedge dx^2 \wedge d\theta \wedge dU
\, ,
\end{eqnarray}
and the period of the compact direction $\theta$ is as following
\begin{eqnarray}
\theta\sim \theta+\frac{\pi R_{AdS}^{'2}}{U_\Lambda}
\, .
\end{eqnarray}
Again, we use the coordinate of equation (15) to expand the metric (36) around $\eta = 0$ just
like what we did in the previous section. Then we find
\begin{eqnarray}
{\rm d}s^2 \sim \left( \frac{U_\Lambda}{R'_{AdS}} \right)^{2} (1 + 2 \eta^2) (
\eta_{\mu\nu} {\rm d} X^\mu {\rm d}X^\nu)  +   R_{AdS}^{'2}
{\rm d} \eta^2 + 4  \left( \frac{U_\Lambda}{R'_{AdS}} \right)^{2} \eta^2  {\rm d} \theta^2 + R_{S^1}^2 d\varphi^2 .
\end{eqnarray}
Then we consider fluctuations in the directions transverse to the wall
and expand the Polyakov action (16) around the solutions (15). By keeping
quadratic terms in $\eta$, we obtain the following action
\begin{eqnarray}
S =
\frac{1}{2\pi\alpha'} \left( \frac{U_\Lambda}{R'_{AdS}} \right)^{2}  \int\!{\rm d}\tau{\rm d}\sigma\,
 \frac{R_{AdS}^{'4}}{U_\Lambda^2} \left[ \left( \dot\eta^2 - {\eta'}^2\right)
 - b'\,\cos^2(\sigma)
\left( 1 + 2 \eta^2 \right) \,\right]
\nonumber\\
+ \left[ \left(1 + 2 \eta^2\right) \left(\delta \dot{X}^\mu \delta
  \dot{X}^\nu \eta_{\mu\nu}  - \delta X^{\mu '} \delta X^{\nu'}
  \eta_{\mu\nu}\right) \right] \, ,
\end{eqnarray}
where the dimensionless parameter $b'$ is
\begin{eqnarray}
b' \equiv \frac{2 L^2 U_\Lambda^2}{R_{AdS}^{'4}} \, .
\end{eqnarray}
In the case of $b' \ll 1$, we have the flat space approximation because
the fluctuations of the string are small. Then we can use the coordinate
\begin{eqnarray}
z = \frac{R_{AdS}^{'2}}{U_\Lambda}\, \eta\, ,
\end{eqnarray}
to write the metric (40) in the following conformally flat form
\begin{eqnarray}
{\rm d}s^2 \sim \left( \frac{U_\Lambda}{R'_{AdS}} \right)^{2}
\Big(\eta_{\mu\nu} {\rm d} X^\mu {\rm d}X^\nu
 + {\rm d}z^2 \Big) + \left( \frac{U_\Lambda}{R'_{AdS}}\right)^{2} \eta^2 {\rm d}\theta^2\,+  R_{S^1}^2 d\varphi^2 ,.
\end{eqnarray}
Again we expand the Polyakov action for the fluctuations in the
transverse directions and use $T = L \tau$, we can find
\begin{eqnarray}
S_{\text{fluct}} = \frac{L}{2 \pi \alpha'_{\text{eff(2)}}} \int\!{\rm d} T {\rm d} \sigma\,
\left[ - (\partial_T z)^2 + \frac{1}{L^2} (\partial_\sigma z)^2\right] \, ,
\end{eqnarray}
where
\begin{eqnarray}
\alpha'_{\text{eff(2)}} = \alpha' \left( \frac{R'_{AdS}}{U_\Lambda} \right)^{2} \, ,
\end{eqnarray}
is the effective string coupling for the near extremal $KM$ model.  Because of the $AdS$ radii differences
in these two non-critical backgrounds,  this equation is not the same as equation (22).\\
Then we put equation (24) for the fluctuations $z(\sigma,\tau)$ into the action (44)
and integrate over $\sigma$ coordinate to find
\begin{eqnarray}
S_{\text{fluct}} = \frac{L}{2 \alpha'_{\text{eff(2)}}} \int\!{\rm d} T\,
    \left[  \sum_{n>0}\left( - (\partial_T z_n)^2 +
		\frac{n^2}{L^2}
  z_n^2\right)\right]\,.
\end{eqnarray}
This action is also similar to the action for an infinite number of linear
harmonic oscillators with frequencies $n/L$ and masses $L/\alpha'_{\text{eff(2)}}$.
Then by using the process of previous section, we find the string
fluctuation probability as following
\begin{eqnarray}
{\mathcal P}^{\text{min}}_{\text{fluct}} \approx \exp\left( - 1.3\frac{z_B^2}{\alpha'_{\text{eff(2)}}}\right) \, .
\end{eqnarray}
By inserting equation (47) into equation (6), we obtain the following equation
for the decay width of large-spin mesons in the near extremal $KM$ model
\begin{eqnarray}
\Gamma_{\text{flat}} =  \Big( \text{const}.\times
T_{\text{eff}}\,{\mathcal P}_{\text{split}}\,\times\,
 L \Big) \,\times\, \exp\left( - 1.3\frac{z_B^2}{\alpha'_{\text{eff(2)}}}\right)  \, ,
\end{eqnarray}
where $z_B$ is the position of the flavor brane. This equation is for the flat space
approximation and is similar to the near extremal $AdS_6$ results and
also ref. [37]. These results are different in the precise form of the exponent which
depend on effective string couplings. This is the decay width obtained
by using the near extremal $KM$ model with the $AdS_5 \times S^1$ black
hole background. We present our numerical results in the last section.

\section{Results and discussion}
In this section, we present the numerical results for the decay widths
and compare them with the models in [37] and the experimental data of ref. [50].\\
\begin{figure}[bth]
\centerline{\includegraphics[width=7cm]{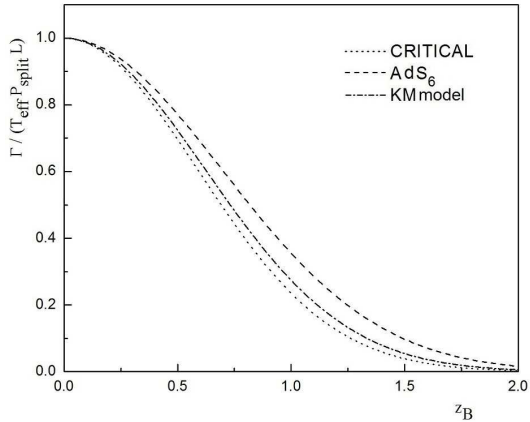}} \caption{ The decay widths of
mesons vs. the position of flavor brane. The dash line belongs to the $AdS_6$ model,
the dash-dot line belongs to the near-extremal KM model, and the dotted line is for
the critical case.The decay width in the non-critical models
are more than the critical one.
\label{fig1}}
\end{figure}

As mentioned before, the difference between equations (34) and (48) and the critical model [37]
is the different effective string couplings.
We use these equations and obtain the decay widths numerically. For this purpose,
we put the values $ R_{AdS} = \sqrt{\frac{15}{2}} $, $R'_{AdS} = \sqrt{6} $ and $T_{\text{eff}}\approx 0.177~\text{GeV}^2$
into the equations (34),(48) and use equations (22),(45) to plot the decay width (fig. 1).
From this diagram, we can see that the decay widths of mesons in the non-critical backgrounds
of sections 2,3 are more than in the critical model. So we find larger values for the decay widths
compared to the critical model.\\
Then we plot $\Gamma /M$ in terms of $M$ for two large-spin mesons, $a$ and $f$
by using a sigmoidal fitting for the experimental data of ref. [50]  (see fig.2 ).
From this diagram, we can see that the decay widths in the nature are not constant
on the same Regge trajectory. The decay widths behave like equation (35) in which
the effect of  two massive endpoints of  the string is considered.\\
\begin{figure}[bth]
\centerline{\includegraphics[width=16cm]{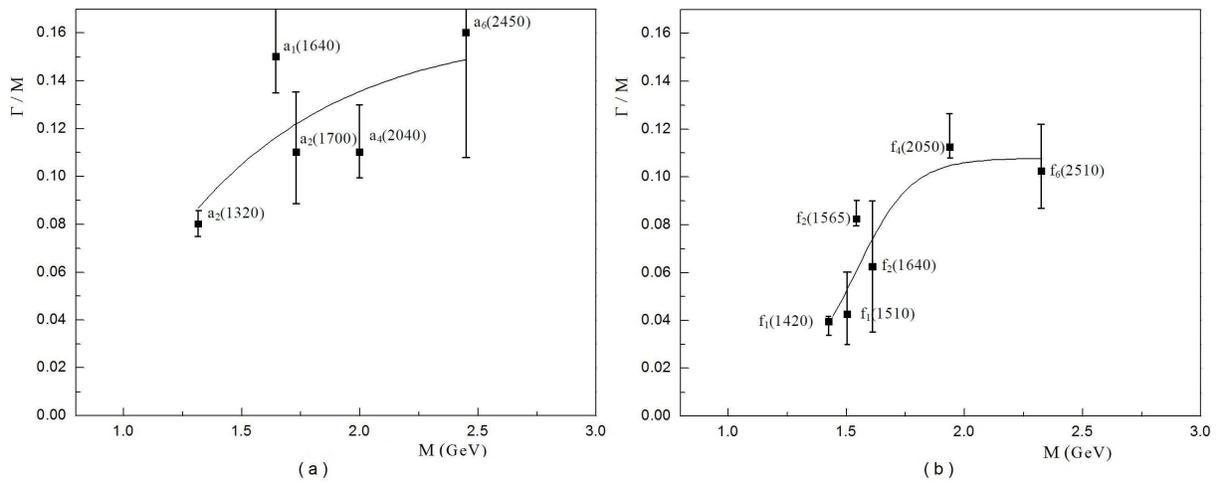}} \caption{a) The decay width divided
by mass versus the mass of the mesons states on the $a$ trajectory.
b) The decay width divided
by mass versus the mass of the mesons states on the $f$ trajectory. The solid lines correspond
to a sigmoidal fitting for the data [50].
\label{fig2}}
\end{figure}

Also we can use equation (35) for the two non-critical models of
sections 2,3 and the critical model [37] to plot  $\Gamma /M$ in terms of $M$ (fig.3).\\
In this figure we can see that for the $a$ trajectory, there is a good agreement between the
$AdS_6$ diagram and the experimental data for all spin values (the left diagram).
For the $f$ trajectory we can see that our results deviate from the experimental results
in low spins and the fitting of experimental data has a better agreement with the critical
model of ref. [37] but for high spins in the $f$ trajectory, the $AdS_5 \times S^1$
diagram is more compatible with the experimental data (the right diagram).\\
\begin{figure}[bth]
\centerline{\includegraphics[width=16cm]{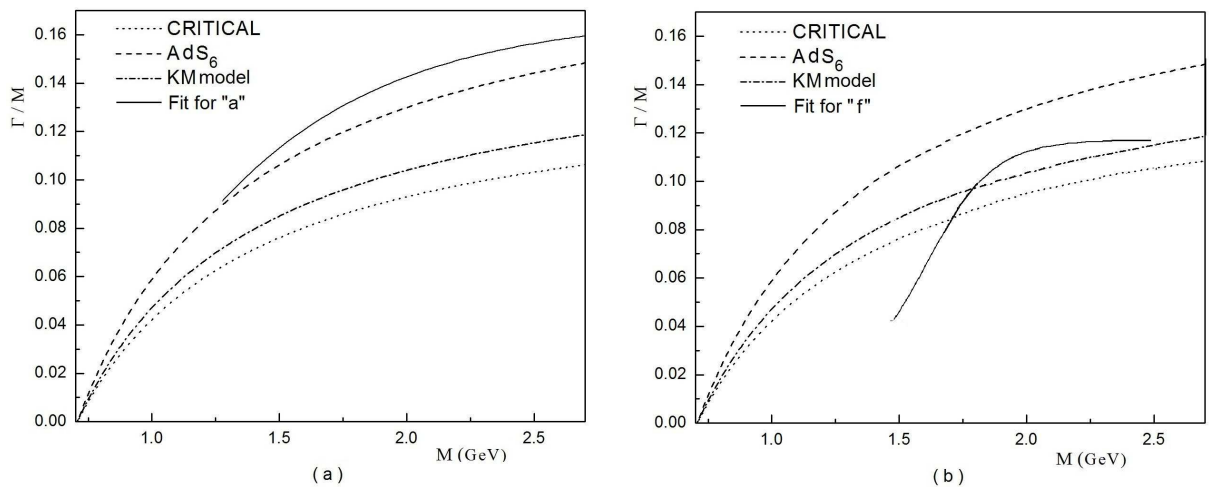}} \caption{a)
Comparison of the $\Gamma /M$ vs. $M$ for two non-critical models,
critical model [37], and the fitting for $a$ meson. The dash line
belongs to the $AdS_6$ model, the dash-dot line belongs to the
$AdS_5\times S^1$ KM model, and the dotted line is for the critical case.
The solid line is the fit for $a$ meson.
b) Comparison of the $\Gamma /M$ vs. $M$ for two non-critical models,
critical model [37], and the fitting for $f$ meson.
\label{fig3}}
\end{figure}\\
In this paper we used the non-critical string/gauge duality and obtained the decay widths for
large-spin mesons. We chose two different six dimensional black hole backgrounds;
the near extremal $AdS_6$ background and the near extremal KM model. By using the method of ref.[37],
we obtained expressions for the decay widths and plotted it in terms of the position of
the flavor brane (fig.1). From this diagram it is easy to see that the non-critical models lead to
larger values for the decay width. Then we used another equation from ref. [39]
and plotted the decay width in terms of the masses of meson states. We compared these results with the
fitting of experimental data of ref. [50] for mesons $a$ and $f$ (fig.3). From these diagrams one can find that
for $a$ meson, the $AdS_6$ background leads to a better result. Also for $f$ meson, the near extremal KM
model has better agreement with data only for large spins. For lower spins, the critical model of ref. [37]
is closer to the experimental data.\\


\end{document}